\documentstyle[aps,12pt]{revtex}

\begin{document}
\draft
\author{V. V. Ulyanov$^{1}$ and O. B. Zaslavskii$^{2}$}
\address{Department of Physics, Kharkov State University, Svobody Sq.4, Kharkov\\
310077, Ukraine\\
E-mail addresses: $^{1}$vladimir.v.ulyanov@univer.kharkov.ua\\
$^{2}$aptm@kharkov.ua}
\title{Spin systems, spin coherent states and quasi-exactly solvable models}
\maketitle

\begin{abstract}
{Spin coherent states play a crucial role in defining QESM (quasi-exactly
solvable models) establishing a strict correspondence between energy spectra
of spin systems and low-lying quantum states for a particle moving in a
potential field of a certain form. Spin coherent states are also used for
finding the Wigner-Kirkwood expansion and quantum corrections to energy
quantization rules. The closed equation which governs dynamics of a quantum
system is obtained in the spin coherent representation directly for
observable quantities.}
\end{abstract}


{Spin coherent states play a crucial role in defining QESM (quasi-exactly
solvable models) establishing a strict correspondence between energy spectra
of spin systems and low-lying quantum states for a particle moving in a
potential field of a certain form. Spin coherent states are also used for
finding the Wigner-Kirkwood expansion and quantum corrections to energy
quantization rules. The closed equation which governs dynamics of a quantum
system is obtained in the spin coherent representation directly for
observable quantities.}

In this paper we review three somewhat unusual applications of spin coherent
states: 1) quasi-exactly-solvable models (QESM) and effective potential
description of spin systems; 2)dynamics of quantum spin systems in terms of 
{\it observable} quantities; 3) Wigner-Kirkwood expansion and energy
quantization rules (analogue of the Bohr-Sommerfeld rules) with quantum
corrections, derivation not using the path integral approach; the crucial
point here that the series for quantization rules turned out to be the
direct consequence of the Wigner-Kirkwood expansion, so the present approach
establishes the connection between two quite different expansions. It is
essential that all three points under discussion which look so different are
based on the possibility to represent the spin operators as the differential
ones.

Let us consider the standard expression for a spin coherent (not normalized)
state: 
\begin{equation}
|\xi \rangle =\exp (\xi S_{-})|S\rangle =\sum_{\sigma =-S}^{S}\sqrt{\frac{%
(2S)!}{(S-\sigma )!(S+\sigma )!}}\xi ^{S-\sigma }|\sigma \rangle  \label{coh}
\end{equation}
where $|\sigma \rangle $ denotes the state with the $S_{z}$ projection equal
to $\sigma $, $S_{\pm }=S_{x}\pm iS_{y}$. Then using the commutation
relation for different projections of spin operators we obtain that for any
function $f$ of spin operators $S_{i}$: 
\begin{equation}
\langle \xi |S_{i}f|\xi \rangle =\ddot{S}_{i}f  \label{aver1}
\end{equation}
where $f=$ $\langle \xi |f|\xi \rangle $ and 
\begin{equation}
\ddot{S}_{+}=\frac{\partial }{\partial \xi ^{*}}\text{, }\ddot{S}_{-}=-\xi
^{*2}\frac{\partial }{\partial \xi ^{*}}\text{, }\ddot{S}_{z}=-\xi ^{*}\frac{%
\partial }{\partial \xi ^{*}}  \label{dif1}
\end{equation}

Another version of the representation of spin operators in terms of
differential ones arises if one uses normalized spin coherent states $|\vec{n%
}\rangle =(1+\xi \xi ^{*})^{-S}|\xi \rangle $. Then, similarly to (\ref
{aver1}) we have 
\begin{equation}
\langle \vec{n}|S_{i}f|\vec{n}\rangle =\hat{S}_{i}f  \label{aver2}
\end{equation}
where now $f=\langle \vec{n}|f|\vec{n}\rangle $. Here $\vec{n}$ is the unit
vector whose direction is parametrized by two angles or a complex number $%
\xi $ according to $\xi =\tan \frac{\theta }{2}\exp (i\phi )$. The explicit
expressions for $\hat{S}_{i}$ are the following: 
\begin{equation}
\hat{S}_{x}=\frac{S(\xi +\xi ^{*})}{1+\xi \xi ^{*}}+\tilde{S}_{x}\text{, }%
\hat{S}_{y}=\frac{S(\xi -\xi ^{*})}{i(1+\xi \xi ^{*})}+\tilde{S}_{y}\text{, }%
\hat{S}_{z}=\frac{S(1-\xi \xi ^{*})}{(1+\xi \xi ^{*})}+\ddot{S}_{z}
\label{difclas}
\end{equation}
These expressions can be rewritten in the form 
\begin{equation}
\vec{S}=S\vec{n}+\frac{1}{2}(\hat{a}-i\hat{b})\text{, }\hat{a}=-\vec{n}%
\times \hat{b}\text{, }\hat{b}=\vec{n}\times \nabla \text{, }\nabla =\frac{%
\partial }{\partial \vec{n}}  \label{difn}
\end{equation}
The formulas for $\ddot{S}_{i}$ and $\hat{S}_{i}$ play the key role in what
follows.

\section{QESM and spin coherent states}

Quasi-exactly solvable models is an rather unusual object in quantum
mechanics which occupies a position intermediate between exactly solvable
models and models which cannot be solved at all. At present, there are
several reviews on QESM \cite{shif}, \cite{rev}, \cite{ush}, \cite{turb1}, 
\cite{turb2} made from different viewpoints where a reader can find
references to original papers and history of discovering QESM. In the
present paper we outline briefly aspects of QESM connected with their
physical realization.

Usually, the typical situation in quantum mechanics with exact solutions of
the Schr\"{o}dinger equation is the following. (1) The expressions for wave
functions and energy levels can be found for a whole spectrum; (2) a hidden
underlying algebraic structure which makes it possible to find exact
solutions has the auxiliary character which in itself has no direct physical
meaning; (3) the possibility to describe some object by a potential field
which admits exact solutions is determined by comparison with an experiment
but not by inner structure of the problem. In contrary, for QESM (1) only
the part of the spectrum can be found explicitly or implicitly from the
algebraic equation of finite degree; (2)-(3) the underlying algebraic
structure (spin Hamiltonian) has direct physical meaning, so potential
description of spin systems arises because of just the spin structure itself
in a rigorous sense; that leads to the notion of an essentially new type of
quasi-particle which can be called ''spinon''.

Let us consider the spin Hamiltonian 
\begin{equation}
H=a_{ij}S_{i}S_{j}+b_{i}S_{i}  \label{ham}
\end{equation}
The representation for spin operators in terms of differential ones (\ref
{dif1}) enables one to obtain for the eigenvalue problem the second order
differential equation which after a simple substitution and, in general, the
change of variables, leads to the standard Schr\"{o}dinger equation with
some potential. Below we discuss several examples. Let, first 
\begin{equation}
H=-S_{z}^{2}-BS_{x}  \label{oneax}
\end{equation}
that describes an uniaxial paramagnet in an transverse magnetic field. Then
in the corresponding Schr\"{o}dinger equation the potential $U=B^{2}/4\sinh
^{2}x$ - $B(S+\frac{1}{2})\cosh x$, the wave function $\Psi =\Phi \exp [-(%
\frac{B}{2}\cosh x)]$ where $\Phi =\sum_{\sigma =-S}^{\sigma =S}a_{\sigma
}e^{\sigma x}$ with some coefficients $a_{\sigma }$. It follows from the
form of the wave function that it decays rapidly at infinity and, therefore,
describes bound states. On the basis of the oscillation theorem it follows
from the form of the wave function that the spin energy levels coincide with
the initial $2S+1$ energy levels of the particle (''spinon'') moving in the
potential under discussion. The higher levels have nothing to do with the
spin system in question. The found effective potential undergoes a curious
transformation as the magnetic field changes. For $B>B_{0}=2S+1$ it has the
form of a single well, for $B<B_{0}$ it changes into a double well, for $%
B=B_{0}$ it takes the form of a well with a fourfold minimum.

Near the critical magnetic field $B=B_{0}$ the potential can be approximated
by a power expansion. and represents, in fact, a quartic ahnahrmonic
oscillator. Using properties of such a system, one can show that for the
paramagnet at hand the magnetic susceptibility has a maximum at $%
B=B_{0}[1-\gamma (S+\frac{1}{2})^{-2/3}]$ where $\gamma \sim 1$. This
maximum does not disappear in the limit $S\rightarrow \infty $ and in this
sense has a pure quantum origin. Another application of the effective
potential method consists in the possibility to calculate tunnelling rates
for $B<B_{0}$ using well known methods of quantum mechanics (WKB,
instantons, etc.). Even much more important is that the effective potential
description gives clear qualitative understanding of what the phenomenon of
spin tunnelling is and in what sense spin, which is a quantum object of pure
discrete nature, can tunnel through classically forbidden region.

It turns out that in general the effective potential describing spin systems
is periodic, spin levels corresponding to edges of energy bands. For
instance, for $H=\alpha S_{z}^{2}-\beta S_{y}^{2}+BS_{x}$ the potential $U$
is expressed in terms of elliptic functions, the condition which select spin
levels reads $\Psi (x+4K)=(-1)^{2S}\Psi (x)$ where $K=K(k)$ is the complete
elliptic integral of the first kind, $k=\sqrt{\beta /(\alpha +\beta )}$.

Sometimes the infinite Hilbert space of a quantum system can be divided to a
set of finite subspaces with respect to the value of some integral of motion 
$R.\,$Then in each subspace one can introduce its own effective potential.
In this sense for the Dicke spin-boson model with $H=\omega
a^{+}a+\varepsilon S_{z}-g(a^{+}S_{-}+S_{+}a)$ we have $%
U=r^{6}-Ar^{4}+Br^{2}+Cr^{-2}$(we do not give the values of constants for
shortness). The similar potential with $C=0$ corresponds to two interacting
oscillators $H=\omega a^{+}a+\Omega b^{+}b+g(a^{+}b^{2}+ab^{+2})$. In
general, QESM demonstrate a lot of nontrivial correspondences between
spectra of quite different quantum systems.

\section{Dynamics of spin systems}

Consider the Heisenberg equation for an arbitrary operator $\hat{g}(\vec{S})$
in the case of a time-independent Hamiltonian $H$ 
\begin{equation}
\hat{g}=(i/\hbar )(\hat{H}\hat{g}-\hat{g}\hat{H})  \label{heis}
\end{equation}
and average it over a spin coherent state. Then, using relations (\ref
{difclas}), (\ref{difn}) we obtain 
\begin{equation}
\dot{g}=(i/\hbar )\hat{K}g\text{, }g=\langle \vec{n}|g|\vec{n}\rangle \text{%
, }\hat{K}=H(\vec{S})-c.c.  \label{dyn}
\end{equation}
This is the closed equation for an arbitrary quantum system. In the
classical limit it turns into the equation $\dot{g}=\left\{ H_{cl},g\right\} 
$ where $\left\{ ...\right\} $ denote the Poisson bracket which contains
derivatives with respect to the component of a classical spin
(magnetization) of the first order only. The equation of motion has the same
form for any quantity and the only point where the distinction between
different solution comes from is the initial condition: $g(t=0)$ should be
specified as a function of $\vec{n}$ (or $\xi $ and $\xi ^{*}$). As a matter
of fact, variables which parametrize a spin coherent state play the role of
quantum generalization of Lagrange (but not Euler) coordinates. It is
remarkable that the equation under discussion is obtained directly in terms
of averages, i.e. observable quantities, so the stages of finding the wave
function and the subsequent averaging are avoided completely.

Consider the following example. Let the Hamiltonian have the form $%
H=-BS_{x}-DS_{x}^{2}$ and $D\ll B/S$. Then one can show that account for
higher derivatives in the Sch\"{o}dinger equation gives rise to a pure
quantum modulation of a classical periodic dependence: $\left\langle
S_{+}\right\rangle =S\sin \theta \exp [(i(\phi -\omega t)][\cos \tau +i\sin
\tau \cos \theta )^{2S-1}$, $\tau =Dt/\hbar $, $\omega =B/\hbar $.

\section{Quasiclassical approximation for spin systems}

Spin is essentially quantum object having a discrete nature. On the other
hand, in the classical limit a spin system is described by the classical
Hamiltonian function in which the role of natural variables is played by two
angles (for each spin), e.g. variables which change continuously. Therefore,
if one is interested in constructing the analogue of the Wigner-Kirkwood
expansion in powers of $S^{-1}$ the following question immediately arises:
how can these two circumstances be reconciled? The ideal tool to handle this
problem is the apparatus of spin coherent states: (1) they ensure continuous
representation of a spin; (2) they minimize the Heisenberg uncertainty
relation, so they are ''the most classical states'' and in this sense are
already adjusted for the description of the quasiclassical limit and finding
quantum corrections; (3) they form complete (even overcomplete set of
states). Using spin coherent states as a basis we can construct the
expansion in question as the perturbation theory with respect to derivatives
according to (\ref{difn}). In particular, the first correction for
one-particle Hamiltonian $H=f(\vec{S})$ turns out to be $\delta F=\frac{1}{4}%
S^{-1}\sum_{k,l}\left\langle (\delta
_{kl}-n_{k}n_{l})(f_{,k,l}-T^{-1}f_{,k}f_{,l})\right\rangle $ where $\delta
_{kl}$ is the Kronecker delta, $f_{,k}=\frac{\partial f}{\partial n_{k}}$,
angular brackets indicate averaging over the classical Gibbs distribution
with the corresponding classical Hamiltonian function $f(S\vec{n})$, $T$ is
a temperature.

It is remarkable that, knowing the Wigner-Kirkwood series, one may recover
from it the form of the energy quantization rules with quantum corrections
without approximate solving the Schr\"{o}dinger equation. For the
''ordinary'' quantum mechanics it was shown in \cite{ulyan} and is extended
now directly to spin systems.

To summarize, spin coherent states not only establish link between quantum
and classical spin systems - they even lead to such constructions which
(like QESM) in themselves have nothing to do with spin!

The work of O. Z. is supported by ISF, grant \# QSU080268.

References





%
%

%
%

\end{document}